\documentclass[12pt]{iopart}

\begin{document}

\title[Quantum operational measurement ...]{Quantum operational
measurement of amplitude and phase parameters for SU(3)-symmetry
optical field}

\author{A.P.Alodjants, A.Yu.Leksin, S.M.Arakelian}

\address{Department of Physics and Applied Mathematics,
Vladimir State University \\ 600000, Vladimir, Russia}

\eads{\mailto{laser@vpti.vladimir.ru}}

\begin{abstract}
We consider a new approach to describe a quantum optical
Bose-system with internal Gell-Mann symmetry by the SU(3)-symmetry
polarization map in Hilbert space. The operational measurement in
density (or coherency) matrix elements for the three mode optical
field is discussed for the first time. We have introduced a set of
operators that describes the quantum measurement procedure and the
behavior of fluctuations for the amplitude and phase
characteristics of three level system. The original twelve-port
interferometer for parallel measurements of the Gell-Mann
parameters is proposed. The quantum properties of W-qutrit states
under the measurement procedure are examined.
\end{abstract}

\submitto{\JOB}

\pacs{42.50.Dv, 03.67.Hk, 42.81Qb}


\section{Introduction}
At present, the properties of three level optical and atomic
systems (qutrit states) evoked a great interest in modern problems
of quantum information and communication [1-5]. In fact,
implementation of qutrit states in quantum system is more
preferable than qubit presentation in some cases - see also [3].
Establishment of qutrit states in quantum optics can be realized
by different ways [2,4,5]. One of them is taking into account the
three optical mode entanglement at multiport beam-splitter
(tritter) [4]. Other possibility is based on exploiting nonlinear
spontaneous down conversion (SPDC) process as a source of
entangled linearly polarized photons [5]. In this case the
polarization characteristics for biphoton state can be represented
by three components of polarization vector.

Another keystone problem is to measure the qutrit state
characteristics. Nowadays the quantum state tomography approach is
being discussed for that [6-8]. The basic idea of the method is to
reconstruct the density matrix elements by a set of measurements
(projections)\textit{.} The quantum tomography for low dimensional
systems, e.g. for spin states, has been firstly proposed in
[9,10]. It has been shown that the density matrix elements can be
reconstructed from marginal distributions - for especially
prepared diagonal elements of density matrix with the help of
unitary transformation. In quantum optics the density (or
coherency) matrix elements can be expressed in terms of SU(2)
observables - Stokes parameters of optical field [11,12]. In fact,
the procedure of spin state tomography is very close to the usual
classical ellipsometric measurement technique [12,13]. For last
case the measured intensity of light $I\left( {\chi ,\delta}
\right) = \frac{{1}}{{2}}\left( {S_{0} + S_{1} \cos\left( {\chi}
\right) + S_{2} \sin\left( {2\chi}  \right)\cos\left( {\delta}
\right) + S_{3} \sin\left( {2\chi}  \right)\sin\left( {\delta}
\right)} \right)$ depends on $\delta $ and $\chi $ parameters
introduced by linear optical elements, i.e., phase plates,
polarizators. Thus, we need at least four measurements of the
Stokes parameters $S_{j} $ ($j = 0,1,2,3$) for complete
determination of the light polarization state. In quantum domain
the Stokes operators $S_{1,2,3} $ do not commute with each other,
and therefore the procedure of the measurement of these parameters
should be clarified. In general, for quantum optical field we are
able to carry out the measurements consecutively with any desired
accuracy. But in this case the number of copies of initial quantum
state of the system is absolutely necessary in order to do the
measurements [6,7].

According to alternative operational approach, the measuring
apparatus operates with initial quantum state of the system and
performs all measurements for non-commuting observables
simultaneously with some accuracy determined by uncertainty
relations. The approach becomes more preferable in some cases, and
results in obtaining unique information about quantum
characteristics of the system, e.g. for the problem of optical
phase measurements [14] and quantum polarization phase
characteristics determination [15]. In particular, in [15] we
consider a special multi-port interferometer for simultaneous
measurements of all polarization Stokes parameters of light. The
quantum error of the measurement is determined by vacuum field
fluctuations, and plays a principal role in the case as it can not
be avoided due to quantum nature of optical field. Such an
approach permits to obtain the complete information about the
density matrix elements (i.e. coherency) and to evaluate the
degree of polarization for light as well.

In this paper we are developing a quantum theory for the
SU(3)-polarization symmetry systems. The approach is based on
three mode interaction in quantum optics in general. We analyze
the quantum phase and amplitude properties in three level system.
In section 2 the mathematical description of the
SU(3)-polarization states for optical fields is presented. Section
3 is devoted to the problem of operational determination of
non-diagonal elements of the coherency matrix and degree of
polarization with the help of the Gell-Mann parameters of optical
field by simultaneous measurement by means of the twelve-port
interferometer. The measurement of amplitude characteristics of
three mode optical field is considered in section 4.

\bigskip

\section{Quantum description of SU(3)-polarization for
Bose-systems}

The quantum three-mode Bose-system with SU(3)-symmetry can be
described in the Schwinger representation by Hermitian Gell-Mann
operators $\lambda _{j} $ ($j = 0,1,...,8$) (cf. [8]):

\numparts
\begin{eqnarray}
\label{eq1111} \lambda _{0} = a_{1}^{ +}  a_{1} + a_{2}^{ +} a_{2}
+ a_{3}^{ +}  a_{3}, \\ \lambda _{1} = a_{1}^{ +}  a_{2} + a_{2}^{
+}  a_{1} , \quad \lambda _{2} = i\left( {a_{2}^{ +}  a_{1} -
a_{1}^{ +}  a_{2}}  \right), \quad \lambda _{3} = a_{1}^{ +} a_{1}
- a_{2}^{ +}  a_{2} , \\ \lambda _{4} = a_{1}^{ +}  a_{3} +
a_{3}^{+}  a_{1} , \quad \lambda _{5} = i\left( {a_{3}^{ +} a_{1}
- a_{1}^{ +}  a_{3}}  \right), \\ \lambda _{6} = a_{2}^{ +}  a_{3}
+ a_{3}^{ +}  a_{2} , \quad \lambda _{7} = i\left( {a_{3}^{ +}
a_{2} - a_{2}^{ +}  a_{3}}  \right),\\ \lambda _{8} =
\frac{{1}}{{\sqrt {3}} }\left( {a_{1}^{ +}  a_{1} + a_{2}^{ + }
a_{2} - 2a_{3}^{ +}  a_{3}}  \right),
\end{eqnarray}
\endnumparts

\noindent where $a_{j} $($a_{j}^{ +}  $), $j = 1,2,3$ are the
photon annihilation (creation) operators. The operators $\lambda
_{j} $ defined above in Eq. (1) obey standard commutation
relations for SU(3)-algebra - see e.g. [8].

In expression (1a) the operator $\lambda _{0} $ determines the
total number of photons; the operators $\lambda _{1,2,3} $ present
the SU(2) sub-group of the SU(3)-algebra. In quantum optics this
sub-group corresponds to the polarization Stokes parameters
$S_{1,2,3} $ for the optical field, where the modes 1 and 2 are
two linear (circular) polarization components of light [11]. The
operators $\lambda _{4,5} $ and $\lambda _{6,7} $ in expressions
(1c,d) characterize coupling between first two modes ($j=1,2$) and
the third one ($j=3$), respectively. The operator $\lambda _{8} $
in Eq. (1e) is described by a combination of photon numbers for
all modes of quantum field.

Let us introduce the unit vector $\vec {e}$ for a three mode
system in Hilbert space:

\begin{equation}
\label{eq6}
\vec {e}a = \vec {e}_{1} a_{1} + \vec {e}_{2} a_{2} + \vec {e}_{3} a_{3}
\end{equation}

\noindent where $a$ is annihilation operator for a three-mode
field; $\vec {e}_{j} $($j = 1,2,3$) are orthogonal vectors under
the condition

\begin{equation}
\label{eq7}
\sum\limits_{j = 1}^{3} {\left| {\vec {e}_{j}}  \right|^{2}} = 1.
\end{equation}

The relation (\ref{eq6}) can be written as:

\begin{equation}
\label{eq8}
a = e_{1}^{\ast}  a_{1} + e_{2}^{\ast}  a_{2} + e_{3}^{\ast}  a_{3} ,
\end{equation}

\noindent where $e_{j}^{\ast}  = \vec {e}^{\ast} \vec {e}_{j} $
are the projections of vector $\vec {e}$.

The expressions (2)--(4) determine decomposition of a three-mode
optical field by the analogy of usual decomposition of
elliptically polarized light in respect of two orthogonal
(linearly or circularly) polarization components in quantum optics
-- cf.~[13]. However only two orthogonal vectors fulfil to
transversality condition for plane waves. Therefore the physical
meaning of $e_j$ parameters in expressions (2)--(4) should be
clarified for each of specific polarization problems. For example,
we also refer here to some problems of quantum optics when
additional (longitudinal) component of polarization is presented
(see e.g. [16]) and our description can be useful.

Thus, we rewrite expression (\ref{eq8}) for the three-mode problem
in terms of the four parameters $\theta $, $\phi $, $\psi _{1} $,
$\psi _{2} $ according to the SU(3) symmetry approach (see e.g.
[8]):

\begin{equation}
\label{eq9} e_{1} = e^{i\psi _{1}} \sin\theta \cos\phi , \quad
e_{2} = e^{i\psi _{2}} \sin\theta \sin\phi , \quad e_{3} =
\cos\theta ,
\end{equation}
where parameters $\theta,\phi\in[0;\pi/2]$ and
$\psi_{1,2}\in[0;2\pi)$.

The parameters $\theta $ and $\phi $ describe amplitude
characteristics of the three level quantum system. The phase
properties of quantum state of optical field with SU(3)-symmetry
are determined by parameters $\psi _{1} $ and $\psi _{2} $
respectively.

Let us consider the SU(3)-polarization state for coherent optical
field:

\begin{equation}
\label{eq10}
a\left| {\alpha}  \right\rangle = \alpha \left| {\alpha}  \right\rangle ,
\quad
a_{j} \left| {\alpha _{j}}  \right\rangle = \alpha _{j} \left| {\alpha _{j}
} \right\rangle ,
\quad
j = 1,2,3
\end{equation}

\noindent where $\left| {\alpha}  \right\rangle = \left| {\alpha
_{1}}  \right\rangle \left| {\alpha _{2}}  \right\rangle \left|
{\alpha _{3}}  \right\rangle $ is the coherent state of the
three-mode field. With the help of expressions (\ref{eq8})-(6) we
obtain:

\begin{equation}
\label{eq11} \alpha = \sum\limits_{j = 1}^{3} {e_{j}^{\ast}
\alpha _{j}}  , \quad \alpha _{j} = e_{j} \alpha , \quad j =
1,2,3.
\end{equation}

In the paper we also examine the following tripartite state that
can be established as (cf. [17]):

\begin{equation}
\label{eq12}
\left| {\Psi}  \right\rangle _{N} = \frac{{1}}{{\sqrt {N!}} }\left( {e_{1}
a_{1}^{ +}  + e_{2} a_{2}^{ +}  + e_{3} a_{3}^{ +} }  \right)^{N}\left| {0}
\right\rangle .
\end{equation}

\noindent where $\left| {0} \right\rangle \equiv \left| {0}
\right\rangle _{1} \left| {0} \right\rangle _{2} \left| {0}
\right\rangle _{3} $ is a vacuum state; $N = \left\langle {\lambda
_{0}}  \right\rangle $ is the total number of particles. For
microscopic limit, when $N = 1$, we have a tripartite entangled
qutrit W-state for a three level system:

\begin{equation}
\label{eq13}
\left| {\Psi}  \right\rangle _{q} = e_{1} \left| {1} \right\rangle _{1}
\left| {0} \right\rangle _{2} \left| {0} \right\rangle _{3} + e_{2} \left|
{0} \right\rangle _{1} \left| {1} \right\rangle _{2} \left| {0}
\right\rangle _{3} + e_{3} \left| {0} \right\rangle _{1} \left| {0}
\right\rangle _{2} \left| {1} \right\rangle _{3}
\end{equation}

The state $\left| {\Psi}  \right\rangle _{q} $ can be produced by
using a tritter [4]. The maximally entangled state is realized for
$e_j=1\left/\sqrt{3}\right.$ ($\theta=\arccos\left( 1
\left/\sqrt{3}\right. \right)$, but for some problems of quantum
information the maximally-entangled qutrits are not optimal - see
e.g. [1]. In the paper we also consider non-symmetric qutrit
W-states that can be obtained from Eq.(\ref{eq13}) for $\theta =
\phi = \pi /4$.

For $e_{i} = 0$ and $e_{j} \ne 0$ ($i,j = 1,2,3$, $i \ne j$) the
qutrit state $\left| {\Psi}  \right\rangle _{q} $ in
Eq.(\ref{eq13}) reduces to one of three qubit states $\left|
{\Psi}  \right\rangle _{ij} $.

With the help of definitions (1) it is easy to obtain the
following relations for the Gell-Mann parameter variances for the
optical field in state $|\Psi\rangle_N$:

\begin{equation}
\label{eq14} \hspace{-1.8cm} {}_N\left\langle \Psi \right|{\left(
{\Delta \lambda _{0}} \right)^{2}} \left| \Psi \right\rangle _{N}
= 0, \quad {}_N\left\langle \Psi \right| {\left( {\Delta \lambda
_{j}} \right)^{2}} \left| \Psi \right\rangle _{N} \le \left\langle
\alpha \right| {\left( {\Delta \lambda _{j}} \right)^{2}} \left|
\alpha \right\rangle, \quad j = 1,...,8
\end{equation}

\noindent where the expressions $\left\langle \alpha \right|
{\left( {\Delta \lambda _{j}} \right)^{2}} \left| \alpha
\right\rangle$ represent the variances of Gell-Mann parameters (1)
for coherent state (\ref{eq10}). The inequality in (\ref{eq14})
characterizes non-classical properties of entangled states
(\ref{eq12}), (\ref{eq13}) when the optical field fluctuations
that correspond to the Gell-Mann parameter variances are
suppressed below the level of fluctuations for coherent states,
i.e. the effect of squeezing occurs.

\section{Degree of polarization; the SU(3)-interferometer}

Let us consider for the first time the problem of degree of
polarization for optical field with SU(3) symmetry.

We start from classical definition for a two-mode optical system.
In particular in the case of stochastic plane waves the degree of
polarization $P_2$ can be represented as -- see e.g.~[13].:
\begin{equation}
\label{eq11p2} P_2=\left(
1-\frac{4\det\left(J_2\right)}{\left(\Tr\left(J_2\right)\right)^2}\right)^{1/2},
\end{equation}
where $J_2$ is coherency matrix of size $2\times 2$. It is
important that $P_2$ parameter could be expressed in terms of
scalar invariants $\Tr\left(J_2\right)$, $\det\left(J_2\right)$
and
$\Tr\left(\left(J_2\right)^2\right)=\left(\Tr\left(J_2\right)\right)^2
- 2\det\left(J_2\right)$ respectively.

Alternatively the degree of polarization $P_2$ defined in
Eq.~(\ref{eq11p2}) can be rewritten in terms of Stokes parameters
$\langle S_j \rangle$ ($j=0,1,2,3$) for two-mode optical field as:
\begin{equation}
\label{eq12p2} P_2=\frac{\left( \langle S_1 \rangle^2+\langle S_2
\rangle^2+\langle S_3 \rangle^2\right)^{1/2}}{\langle S_0 \rangle}
\end{equation}

Although there are various definitions of degree of polarization
in quantum optics (we do not discuss them in the paper -- see e.g.
[18--20]) the expression~(\ref{eq12p2}) can be used in quantum
domain as well. In this case the $\langle S_j \rangle$ variables
in Eq.~(\ref{eq12p2}) represent expectation values of the Stokes
operators $S_j$.

Now we switch our attention to the case of three-mode optical
system (non-plane waves -- cf.~[13]).

Quantum properties of optical field with SU(3)-symmetry can be
described in this case by the following density (or coherency)
matrix $J_3$:

\begin{equation}
\label{eq15} J_3 = \left( {{\begin{array}{*{20}c}
 {\left\langle {a_{1}^{ +}  a_{1}}  \right\rangle}  \hfill & {\left\langle
{a_{1}^{ +}  a_{2}}  \right\rangle}  \hfill & {\left\langle {a_{1}^{ +}
a_{3}}  \right\rangle}  \hfill \\
 {\left\langle {a_{2}^{ +}  a_{1}}  \right\rangle}  \hfill & {\left\langle
{a_{2}^{ +}  a_{2}}  \right\rangle}  \hfill & {\left\langle {a_{2}^{ +}
a_{3}}  \right\rangle}  \hfill \\
 {\left\langle {a_{3}^{ +}  a_{1}}  \right\rangle}  \hfill & {\left\langle
{a_{3}^{ +}  a_{2}}  \right\rangle}  \hfill & {\left\langle {a_{3}^{ +}
a_{3}}  \right\rangle}  \hfill \\
\end{array}} } \right)
\end{equation}

In general, scalar invariants $\Tr\left( {J_3^{3}} \right)$ and
$\Tr\left( {J_3^{2}} \right)$ take place in this case. In
particular they can be expressed as:
\begin{equation}
\label{eq12new} \hspace{-2cm} \Tr\left( {J_3^{3}}
\right)=P_3^2\left( \Tr\left(J_3\right)\right)^3 + 3 \det\left(
J_3 \right), \qquad \Tr\left( J_3^2\right)=\frac{\left( \Tr\left(
J_3\right)\right)^2}{3}\left(1+2P_3^2\right),
\end{equation}
where $P_3$ represents the degree of polarization for a three-mode
optical field. With the help of the definition of Gell-Mann
operators (1) the $P_{3} $ quantity can be represented as:
\begin{equation}
\label{eq1212} P_{3} = \frac{{\sqrt {3}} }{{2}}\frac{{\left(
{\sum\limits_{j = 1}^{8} {\left\langle {\lambda _{j}}
\right\rangle ^{2}}} \right)^{1/2}}}{{\left\langle {\lambda _{0}}
\right\rangle} }.
\end{equation}

Note that expression~(\ref{eq1212}) for the degree of polarization
$P_3$ for a three-mode optical system can be also considered as
generalization of the well known definition of polarization degree
$P_2$ in Eq.~(\ref{eq12p2}).

The quantum system with SU(3)-symmetry is completely polarized if
and only if:

\begin{equation}
\label{eq18} \det\left( {J_3} \right) = 0, \quad P_{3} = 1.
\end{equation}

It is easy to check that the states (6), (\ref{eq12}) and (9) are
fulfilled conditions (\ref{eq18}).

Let us consider the procedure of measurement of non-diagonal
matrix elements of $J_3$ (see Eq.~(\ref{eq15})). Schematic set-up
for measurement of all phase dependent Gell-Mann parameters is
shown in Fig.1. At the input of the system we have three modes
$b_j$ ($j=1,2,3$). The boxes $B_{j} $ represent the balanced beam
splitters or symmetric cloning machines (see e.g. [21]) to produce
the three modes $a_{j} $($j = 1,2,3$) and their copies (clones)
${a}'_{j} $ at the output. Then, each of the three mode sets is
transformed in two physically identical star-like interferometers
denoted as $I_{1} $ and $I_{2} $ respectively. Thus, we obtain six
Gell-Mann parameters at the output of the device in Fig.1.

Let us precisely analyze the SU(3) twelve-port interferometer
scheme (Fig.2) for operational (simultaneous) measurement of the
three Gell-Mann parameters $\lambda _{j}$. The $100\% $ efficiency
detectors $D$ are the devices for measurement of the photon
numbers $N_{ij} = d_{ij}^{ +}  d_{ij}$, where $d_{ij}$
($d_{ij}^+$) are annihilation (creation) operators for the modes
at the output of the interferometer. They can be represented as
linear combination of input fields $a_j$ and vacuum modes $V_j$
after beam splitters (BS) in Fig.2. The measurement procedure
results in the detection of photon number differences
($N_{ij}^{\left( { -} \right)} i,j = 1,2,3,\quad i < j$):\numparts
\begin{eqnarray}
\label{eq1919} N_{12}^{\left( { -}  \right)} = N_{12} - N_{21} =
\frac{{1}}{{2}}\lambda _{12}^{} + M_{12}, \\ N_{13}^{\left( { -}
\right)} = N_{13} - N_{31} = \frac{{1}}{{2}}\lambda _{13}^{} +
M_{13},\\ N_{23}^{\left( { -}  \right)} = N_{32} - N_{23} =
\frac{{1}}{{2}}\lambda _{23}^{} + M_{23}
\end{eqnarray}
\endnumparts

\noindent where Gell-Mann parameters $\lambda _{ij}^{} $ are
represented as (cf. Eqs.(1)):\numparts
\begin{eqnarray}
\label{eq2020} \lambda _{12}^{} = a_{1}^{ +}  a_{2} e^{i\phi _{2}}
+ a_{2}^{ +}  a_{1} e^{ - i\phi _{2}} , \\ \lambda _{13}^{} =
a_{1}^{ +}  a_{3} e^{ - i\phi _{1}}  + a_{3}^{ +}  a_{1} e^{i\phi
_{1}} , \\ \lambda _{23}^{} = a_{2}^{ +}  a_{3} e^{i\phi _{3}} +
a_{3}^{ +}  a_{2} e^{ - i\phi _{3}}  \quad .
\end{eqnarray}
\endnumparts

The normally ordered operators $M_{ij} $ in Eq.~(17) are
proportional to the operators $a_{j} $, ($a_{j}^{ +}  $) and
vacuum modes $V_{j} $ , $j = 1,2,3$ at the input of the
interferometer. The average values of these operators fulfil the
condition $\langle M_{ij} \rangle = 0$. From Eqs. (18) it is easy
to see that with the phase shifts $\phi _{j} = 0$ for
interferometer $I_{1} $ we measure $\lambda _{1} $, $\lambda _{4}
$ and $\lambda _{6} $ Gell-Mann parameters - Eqs. (1), and if
$\phi _{1} = \frac{{\pi} }{{2}}$, $\phi _{2,3} = - \frac{{\pi}
}{{2}}$ the measurement $\lambda _{2} $, $\lambda _{5} $ and
$\lambda _{7} $ is realized by interferometer $I_{2}$ - see Fig.1.

From Eqs.(17) for the average values of the photon number
difference $\left\langle {N_{ij}^{\left( { -}  \right)}}
\right\rangle $ and for variances $\left\langle {\left( {\Delta
N_{ij}^{\left( { -}  \right)}} \right)^{2}} \right\rangle $ we
obtain:\numparts
\begin{eqnarray}
\label{eq2121}  \hspace{-2cm} \eqalign{ \left\langle
{N_{ij}^{\left( { -}  \right)}} \right\rangle = \frac{{1}}{{2}}
\left\langle {\lambda _{ij}} \right\rangle ,} \\ \hspace{-2cm}
\eqalign{\left\langle {\left( {\Delta N_{ij}^{\left( { -}
\right)}} \right)^{2}} \right\rangle = \frac{{1}}{{4}}\left\langle
{\left( {\Delta \lambda _{ij}^{} } \right)^{2}} \right\rangle +
\frac{{1}}{{4}}\left( {\left\langle {a_{i}^{ +}  a_{i}}
\right\rangle + \left\langle {a_{j}^{ +} a_{j}}  \right\rangle }
\right), \quad i,j = 1,2,3, \quad i < j}
\end{eqnarray}\endnumparts

\noindent
where the condition $\left\langle {M_{ij}}  \right\rangle = 0$ is taken into
account.

The last two terms in Eq.(19b) are determined by vacuum
fluctuations contribution of the modes $V_{j} $ and characterize
the lowest possible level of the considered variances when
$\left\langle {\left( {\Delta \lambda _{ij}^{}}  \right)^{2}}
\right\rangle = 0$.

Let consider an ultimate case for Fig.2. We assume $a_{3}$-mode to
be the control field in coherent state $\left| {\alpha _{3}}
\right\rangle $ with complex amplitude $\alpha _{3} = \left|
{\alpha _{3}}  \right|e^{i\varphi} $ ($\varphi $ is the phase).
The measured mean photon number differences $\left\langle
{N_{ij}^{\left( { -}  \right)}}  \right\rangle $ are: \numparts
\begin{eqnarray}
\label{eq2424} \left\langle {N_{12}^{\left( { -}  \right)}}
\right\rangle = \frac{{1}}{{2}}\left( {\left\langle {\lambda _{1}}
\right\rangle \cos\left( {\phi _{2}}  \right) - \left\langle
{\lambda _{2}}  \right\rangle \sin\left( {\phi _{2}}  \right)}
\right),\\ \left\langle {N_{13}^{\left( { -}  \right)}}
\right\rangle = \frac{{1}}{{2}}\left| {\alpha _{3}}  \right|\left(
{\left\langle {q_{1}} \right\rangle \cos\left( {\phi _{1}} \right)
- \left\langle {p_{1}} \right\rangle \sin\left( {\phi _{1}}
\right)} \right), \\ \left\langle {N_{23}^{\left( { -}  \right)}}
\right\rangle = \frac{{1}}{{2}}\left| {\alpha _{3}}  \right|\left(
{\left\langle {q_{2}} \right\rangle \cos\left( {\phi _{3}} \right)
+ \left\langle {p_{2}} \right\rangle \sin\left( {\phi _{3}}
\right)} \right)
\end{eqnarray}
\endnumparts

\noindent
where $q_{j} = a_{j} e^{ - i\varphi}  + a_{j}^{ +}  e^{i\varphi} $ , $p_{j}
= i\left( {a_{j}^{ +}  e^{i\varphi}  - a_{j} e^{ - i\varphi} } \right)$, ($j
= 1,2$) are the Hermitian quadratures for two other modes.

Let us briefly discuss the properties of polarization degree
$P_{3} $ in Eq.(15) in the case under consideration. In the limit
of the weak control field $\left| {\alpha _{3}}  \right|^{2} \ll
1$ we can measure quantum polarization properties for the two-mode
optical field (for bipartite qutrit state as well - cf.[5]) by
means of the scheme in Fig.2. Note that the relation (15) for
$P_{3} $ in this case will differ from the usual definition of
degree polarization for two-mode system -- see Eq.~(12) and [11].
In the other limit when $\left| {\alpha _{3}}  \right|^{2} \gg
\left\langle {a_{1,2}^{ +}  a_{1,2}} \right\rangle $ the control
field $a_{3} $ plays the role of strong local oscillator field for
simultaneous homodyne measurement of quadratures of two-mode
optical system according to Eqs. (20b,c). In this case we have
$P_{3} \simeq 1$ from Eq.~(15).

\section{Measurements of the $\theta $ and $\phi$ parameters}

In this section we consider the problem of simultaneous
measurement for diagonal elements of matrix $J_3$
(Eq.~(\ref{eq15})). Namely, we focus our attention on the
procedure of measurements for the $\theta $ and $\phi $ amplitude
parameters.

In classical optics sine and cosine of $\theta$ and $\phi$
amplitude parameters (see Eqs.~(\ref{eq9})) can be obtained from
simultaneous measurement of the light intensities
$I_j\sim\left\langle a_j^+a_j\right\rangle$ ($j=1,2,3$) of three
mode optical field.

Now we define $S_{\phi ,\theta}  $~-- and $C_{\phi ,\theta}
$~--~operators in quantum domain for the $\theta $ and $\phi $
parameters:
\begin{equation}
\label{eq27} \eqalign{S_{\phi}  = \sqrt {\frac{{\hat {n}_{2}}
}{{\hat {n}_{1} + \hat {n}_{2}} }} , \qquad\quad C_{\phi}  = \sqrt
{\frac{{\hat {n}_{1}} }{{\hat {n}_{1} + \hat {n}_{2}} }}, \\
S_{\theta}  = \sqrt {\frac{{\hat {n}_{1} + \hat {n}_{2}} }{{\hat
{n}_{1} + \hat {n}_{2} + \hat {n}_{3}} }} , \quad C_{\theta}  =
\sqrt {\frac{{\hat {n}_{3}} }{{\hat {n}_{1} + \hat {n}_{2} + \hat
{n}_{3}} }} ,}
\end{equation}

\noindent where $\hat {n}_{j} = a_{j}^{ +}  a_{j} $ is the photon
number operator and we use Eq.(7) as well. To calculate the
expectation values and variances for amplitude operators we
represent the operators $\hat {n}_{j} $ as $\hat {n}_{j} = \bar
{n}_{j} + \Delta \hat {n}_{j} $ ($j = 1,2,3$), where $\bar {n}_{j}
\equiv \left\langle {\hat {n}_{j}}  \right\rangle $ is the mean
(classical) value of the j-th photon number, $\Delta \hat {n}_{j}
$ is the small fluctuational part of the corresponding operator
with the properties$\left\langle {\Delta \hat {n}_{j}}
\right\rangle = 0$,$\left\langle {\Delta \hat {n}_{i} \Delta \hat
{n}_{j}}  \right\rangle = 0$ ($i,j = 1,2,3, i \ne j$).

After some mathematical calculations for variances of detected
relative amplitudes $\left\langle {\left( {\Delta S_{\phi,\theta}}
\right)^{2}} \right\rangle $, $\left\langle {\left( {\Delta
C_{\phi,\theta}} \right)^{2}} \right\rangle $ we obtain:\numparts
\begin{eqnarray}
\label{eq2929} \left\langle {\left( {\Delta S_{\phi} }
\right)^{2}} \right\rangle = \frac{{\bar {n}_{1}} }{{4\left( {\bar
{n}_{1} + \bar {n}_{2}} \right)^{3}}}\left( {\frac{{\bar {n}_{2}}
}{{\bar {n}_{1}} }\sigma _{1}^{2} + \frac{{\bar {n}_{1}} }{{\bar
{n}_{2}^{}} }\sigma _{2}^{2}}  \right),\\ \left\langle {\left(
{\Delta C_{\phi} }  \right)^{2}} \right\rangle = \frac{{\bar
{n}_{2}} }{{4\left( {\bar {n}_{1} + \bar {n}_{2}}
\right)^{3}}}\left( {\frac{{\bar {n}_{2}} }{{\bar {n}_{1}} }\sigma
_{1}^{2} + \frac{{\bar {n}_{1}} }{{\bar {n}_{2}^{}} }\sigma
_{2}^{2}}  \right),\\ \left\langle {\left( {\Delta S_{\theta} }
\right)^{2}} \right\rangle = \frac{{1}}{{4N^{3}}}\left(
{\frac{{\bar {n}^{2}_{3}} }{{\left( {\bar {n}_{1} + \bar {n}_{2}}
\right)}}\left( {\sigma _{1}^{2} + \sigma _{2}^{2}} \right) +
\left( {\bar {n}_{1} + \bar {n}_{2}}  \right)\sigma _{3}^{2}}
\right), \\ \left\langle {\left( {\Delta C_{\theta} } \right)^{2}}
\right\rangle = \frac{{1}}{{4N^{3}}}\left( {\bar {n}_{3} \left(
{\sigma _{1}^{2} + \sigma _{2}^{2}}  \right) + \frac{{\left( {\bar
{n}_{1} + \bar {n}_{2}} \right)^{2}}}{{\bar {n}_{3}} }\sigma
_{3}^{2}}  \right)
\end{eqnarray}
\endnumparts

\noindent where $N = \bar {n}_{1} + \bar {n}_{2} + \bar {n}_{3} $
is the total average number of photons. In relations (22) the
variances in the photon number $\sigma _{j}^{2} \equiv
\left\langle {\left( {\Delta \hat {n}_{j}} \right)^{2}}
\right\rangle $ ($j = 1,2,3$) characterize the additional terms to
the mean values of the amplitude parameters in the quantum domain.
In a quasi-classical limit (when $N \gg 1$) from Eqs.(22) we have:

\begin{equation}
\label{eq31}
\left\langle {\left( {\Delta S_{j}}  \right)^{2}} \right\rangle \simeq
\left\langle {\left( {\Delta C_{j}}  \right)^{2}} \right\rangle \simeq 0,
\quad
j = \phi ,\theta ,
\end{equation}

\noindent i.e. the role of fluctuations is not essential. For the
three mode Fock state we have $\sigma _{j}^{2} = 0$ and the
conditions (\ref{eq31}) are satisfied as well. The minimal (zero)
value of the relative amplitude fluctuations can be also achieved
for qubit states when one of the parameters $e_{j} = 0$.

The Eqs. (22) are violated for a small average number of photons
($\bar {n}_{j} \ll 1$). In the case of the photon counting
measurement the average value of some amlitude operator $f\left(
{\left\{ {\hat {n}} \right\}} \right)$ is represented as:

\begin{equation}
\label{eq32}
\left\langle {f\left( {\left\{ {\hat {n}} \right\}} \right)} \right\rangle =
\sum\limits_{\left\{ {n} \right\}} {f\left( {\left\{ {n} \right\}} \right)}
W\left( {\left\{ {n} \right\}} \right)
\end{equation}

\noindent
where $W\left( {\left\{ {n} \right\}} \right) \equiv W\left( {n_{1} ,n_{2}
,n_{3}}  \right)$ is joint probability of the set of eigenvalues $n_{1}
,n_{2} ,n_{3} $ that have the form:

\begin{equation}
\label{eq33}
W\left( {\left\{ {n} \right\}} \right) = \left\langle {:\prod\limits_{j =
1}^{3} {\frac{{\left( {a_{j}^{ +}  a_{j}}  \right)^{n_{j}} e^{ - a_{j}^{ +}
a_{j}} }}{{n_{j} !}}} :} \right\rangle
\end{equation}

For weak field ($\bar {n}_{j} \ll 1$ ) the term with $n_{1} =
n_{2} = n_{3} = 0$ should be discarded in Eq.(\ref{eq32}) and the
sum in relation (\ref{eq32}) should be renormalized to the
expression $1 - \left\langle {:\exp\left( { - \sum\limits_{j}
{a_{j}^{ +}  a_{j}} }  \right):} \right\rangle $ - cf. [14].
Finally, for the variances $\left\langle {\left( {\Delta S_{j}}
\right)^{2}} \right\rangle $, $\left\langle {\left( {\Delta C_{j}}
\right)^{2}} \right\rangle $ we obtain:

\begin{equation}
\label{eq34} \hspace{-1.8cm} \eqalign{\left\langle {\left( {\Delta
S_{\phi} } \right)^{2}} \right\rangle = \left\langle {\left(
{\Delta C_{\phi} }  \right)^{2}} \right\rangle =
\frac{{1}}{{4}}\sin^{2}\left( {2\phi}  \right), \quad \left\langle
{\left( {\Delta S_{\theta} } \right)^{2}} \right\rangle =
\left\langle {\left( {\Delta C_{\theta} }  \right)^{2}}
\right\rangle = \frac{{1}}{{4}}\sin^{2}\left( {2\theta}  \right)}
\end{equation}

The minimal (i.e. zero) value of the fluctuations corresponds in
Eqs. (26) to qubit state when some $\bar {n}_{j} = 0$ (but not the
denominator). The maximal value of the variances is $\left\langle
{\left( {\Delta S_{j}} \right)^{2}} \right\rangle = \left\langle
{\left( {\Delta C_{j}} \right)^{2}} \right\rangle =
\frac{{1}}{{4}}$ for non-symmetric qutrits ($\theta = \frac{{\pi}
}{{4}}, \quad \phi = \frac{{\pi} }{{4}}$). The variances are
$\left\langle {\left( {\Delta S_{\theta} }  \right)^{2}}
\right\rangle = \left\langle {\left( {\Delta C_{\theta} }
\right)^{2}} \right\rangle \approx 0.22$ for maximally entangled
qutrit state (9) when $\bar {n}_{1} = \bar {n}_{2} = \bar {n}_{3}
$.

\section{Conclusion}

In the paper we presented the quantum theory of SU(3)-polarization
for a three mode optical field. In particular we define the degree
of polarization that can be connected with scalar invariants in
respect of unitary transformations. We offer special twelve-port
interferometer for operational reconstruction of non-diagonal
elements of coherency matrix $J_3$ and for the measurement of
polarization degree $P_3$ as well. It is important that the scheme
in Fig.~2 contains only linear optical elemants -- balanced beam
splitters -- as input ports for splitting the state for each input
mode $a_{j} $.

To describe amplitude properties of optical field with SU(3)
symmetry we introduce special variables ($S_j$~-- and
$C_j$~--~operators, $j=\theta,\phi$) that can be measured by
direct simultaneous photodetection procedure. We have shown that
qutrit states are characterized by non-vanishing variances of
these variables.

In this paper we do not present the measurement procedure for the
phases $\psi _{1} $ and $\psi _{2} $ - see Eq. (\ref{eq9}). The
operators for the phase parameters $\psi _{1,2} $ can be
introduced operationally by using the interferometer in Fig.2 as
well -- cf.[15].

\ack

One of us (A.P.A) is grateful to Dr. Olga V. Man'ko for fruitful
discussions and remarks. This work is supported by the Russian
Foundation for Basic Research (grants No 01-02-17478 and
05-02-16576) and some Federal programs of the Russian Ministry of
Science and Technology. One of the authors (A.P. Alodjants) is
grateful to the Russian private Found ``Dynasty'' and
International Center for Theoretical Physics in Moscow for the
considerable financial support.

\Figures

\Figure{Schematic set-up for simultaneous measurement of all phase
depending Gell-Mann parameters. Boxes $B_{1} $, $B_{2} $ and
$B_{3} $ transform three input optical modes $b_1$, $b_2$ and
$b_3$ into six modes $a_{j} $ and ${a}'_{j} $ ($j = 1,2,3$)
respectively; $V_{j0}$ are vacuum modes at the input of boxes
$B_j$. Two other boxes $I_{1}^{} $ and $I_{2} $ correspond to two
interferometers for simultaneous measurement of three parameters
$\lambda _{j} $ at the output ports.}

\Figure{The SU(3)-interferometer for simultaneous measurement of
the phase depending Gell-Mann parameters and the phases $\psi
_{1,2} $, $\psi _{1} - \psi _{2} $. The input (output) modes are
denoted as $a_{j} $ ($d_{ij} $) , ($i,j = 1,2,3,\quad i \ne j$).
Modes $V_{j} $ at the input ports are in vacuum states; BS's are
balanced beam splitters, $\phi _{j} $ are phase shifts.}

\end{document}